\documentclass[letterpaper]{article}
\usepackage{aaai23}
\usepackage{times}
\usepackage{helvet}
\usepackage{courier}
\usepackage[hyphens]{url}
\usepackage{graphicx}
\usepackage{natbib}
\usepackage{caption}
\usepackage{amsthm}
\usepackage{amsmath}
\usepackage{bm}
\usepackage{xcolor}
\usepackage{multirow}

\usepackage[colorlinks=true,linkcolor=blue,citecolor=blue,urlcolor=blue]{hyperref}

\title{Deep Reinforcement Learning for Optimal Portfolio Allocation:\\ A Comparative Study with Mean-Variance Optimization}
\author{
    Srijan Sood, \textsuperscript{\rm 1}
    Kassiani Papasotiriou, \textsuperscript{\rm 1}
    Marius Vaiciulis,\textsuperscript{\rm 2}
    Tucker Balch \textsuperscript{\rm 1}
    }
\affiliations{
    \textsuperscript{\rm 1} J.P. Morgan AI Research\\
    \textsuperscript{\rm 2} J.P. Morgan Global Equities; Oxford-Man Institute of Quantitative Finance\\
    \{srijan.sood, kassiani.papasotiriou, marius.vaiciulis, tucker.balch\}@jpmorgan.com
}

\begin{document}

\maketitle

\begin{abstract}
Portfolio Management is the process of overseeing a group of investments, referred to as a portfolio, with the objective of achieving predetermined investment goals. Portfolio optimization is a key component that involves allocating the portfolio assets so as to maximize returns while minimizing risk taken. It is typically carried out by financial professionals who use a combination of quantitative techniques and investment expertise to make decisions about the portfolio allocation.

Recent applications of Deep Reinforcement Learning (DRL) have shown promising results when used to optimize portfolio allocation by training model-free agents on historical market data. Many of these methods compare their results against basic benchmarks or other state-of-the-art DRL agents but often fail to compare their performance against traditional methods used by financial professionals in practical settings.
One of the most commonly used methods for this task is Mean-Variance Portfolio Optimization (MVO), which uses historical time series information to estimate expected asset returns and covariances, which are then used to optimize for an investment objective.

Our work is a thorough comparison between model-free DRL and MVO for optimal portfolio allocation. We detail the specifics of how to make DRL for portfolio optimization work in practice, also noting the adjustments needed for MVO. Backtest results demonstrate strong performance of the DRL agent across many metrics, including Sharpe ratio, maximum drawdowns, and absolute returns.\renewcommand{\thefootnote}{\fnsymbol{footnote}}\footnote{Published at the FinPlan'23 Workshop, the 33rd International Conference on Automated Planning and Scheduling (ICAPS).}\renewcommand{\thefootnote}{\arabic{footnote}}\setcounter{footnote}{0}
\end{abstract}

\section{Introduction}
Portfolio management is a key issue in the financial services domain. It involves allocating funds across a variety of assets, typically to generate uncorrelated returns while minimizing risk and operational costs. Portfolios can constitute holdings across asset classes (cash, bonds, equities, etc.), or can also be optimized within a specific asset class (e.g., picking the appropriate composition of stocks for an equity portfolio). Investors may choose to optimize for various performance criteria, often centered around maximizing portfolio returns relative to risk taken. Since the advent of Modern Portfolio Theory~\cite{markowitz}, a lot of progress has been made in both theoretical and applied aspects of portfolio optimization.
These range from improvements in the optimization process, to the framing of additional constraints that might be desirable to rational investors~\cite{cornuejols2006optimization, li2014online, kalayci2017review, ghahtarani2022robust}. Recently, the community has tapped the many advancements in Machine Learning (ML) to aid with feature selection, forecasting and estimation of asset means and covariances, as well as using gradient-based methods for optimization.

Concurrently, the past decade has witnessed the success of Reinforcement Learning (RL) in the fields of gaming, robotics, natural language processing, etc.~\cite{silver2017mastering, nguyen2019review, su2016line}. The sequential decision making nature of Deep RL, along with its success in applied settings, has captured the attention of the finance research community.
In particular, some of the most popular areas of focus of the application of DRL in finance have been on automated stock trading~\cite{yang2020deep, theate2021application, zhang2020deep, wu2020adaptive}, risk management through deep hedging~\cite{buehler2019deep,du2020deep,cao2021deep,benhamou2020time} and portfolio optimization.
In the following section, we examine the landscape of DRL in portfolio optimization and trading problems. While these approaches exhibit improved performance over previous studies, they do have some shortcomings. For instance, some generate discrete asset trading signals which limit their use in broader portfolio management. Additionally, the majority of these approaches compare results against ML or buy-and-hold baselines, and do not consider classical portfolio optimization techniques, such as Mean-Variance Optimization.

In our work, we aim to compare a simple and robust DRL framework, that was designed around risk-adjusted returns, with one of the traditional finance methods for portfolio optimization, MVO. We train policy-gradient-based agents on a multi-asset trading environment that simulates the US Equities market (using market data replay), and create observation states derived from the observed asset prices. The agents optimize for risk-adjusted returns, similar to the traditional MVO methods. We compare the performance of the DRL strategy against MVO through a series of systematic backtests, and observe improved performance along many performance metrics, including risk-adjusted returns, max drawdown, and portfolio turnover.

\section{Related Work}
There is a lot of recent research interest into the application of Deep RL in trading and portfolio management problems.
For portfolio optimization, a lot of the research focuses on defining various policy network configurations and reports results that outperform various traditional baseline methods~\cite{wang2019alphastock, liang2018adversarial, lu2017agent,jiang2017cryptocurrency,Wang_Huang_Tu_Zhang_Xu_2021, deng2016deep, cong2021alphaportfolio}. Other work explores frameworks that inject information in the RL agent's state by incorporating asset endogenous information such as technical indicators~\cite{liu2020finrl, sun2021deepscalper,du2020stock} as well as exogenous information such as information extracted from news data~\cite{ye2020reinforcement, lima2021intelligent}.

The current benchmarks for DRL frameworks typically involve comparing results against other DRL or ML approaches, a buy-and-hold baseline, or market/index performance. However, these benchmarks may be overly simplistic or provide only a relative comparison. To truly gauge the effectiveness of a DRL agent, it would be more meaningful to benchmark it against methodologies used by financial professionals in practice, such as Mean Variance Optimization (MVO).

While there are some approaches that compare DRL performance with MVO~\cite{li2019optimistic,koratamaddi2021market, i2020deep}, the comparison simply serves as another baseline, and the methodology is not clearly described because an in-depth comparison is not the primary focus of their study.
To our knowledge, there is only one study that goes into a robust in-depth comparison of MVO and DRL~\cite{benhamou2020bridging}.
However, across all these studies, there is usually a discrepancy between the reward function used to train the RL agent, and the objective function used for MVO (e.g., daily returns maximization vs risk minimization). In order to make a fair comparison, it is crucial that both approaches optimize for the same goal. Additionally, some of these approaches provide exogenous information (e.g., signals from news data) to the DRL agent, which makes for a biased comparison with MVO. Moreover, none of these works provide implementation details for the MVO frameworks they used for their comparison. We aim to address these issues by conducting a robust comparison of Deep RL and Mean-Variance Optimization for the Portfolio Allocation problem.

\section{Background}\label{sec:background}

The goal of portfolio optimization is to continuously diversify and reallocate funds across assets with the objective of maximizing realized rewards while simultaneously restraining the risk.
In practice, portfolio management often aims to not only maximize risk-adjusted returns but also to perform as consistently as possible over a given time interval (e.g. on a quarterly or yearly basis).

Markowitz introduced the modern portfolio theory (MPT)~\cite{markowitz}, a framework that allows an investor to mathematically balance risk tolerance and return expectations to obtain efficiently diversified portfolios. This framework relies on the assumption that a rational investor will prefer a portfolio with less risk for a specified level of return and concludes that risk can be reduced by diversifying a portfolio.
In this section, we will introduce Mean Variance Optimization (MVO) -- one of the main techniques of the MPT -- which we later compare to the performance of our DRL framework. Additionally, we introduce RL preliminaries, describing the technique independent of portfolio optimization.

\subsection{Mean-Variance Portfolio Optimization}\label{sec:mvo}
Mean-Variance Optimization (MVO) is the mathematical process of allocating capital across a portfolio of assets (optimizing portfolio weights) to achieve a desired investment goal, usually: 1. Maximize returns for a given level of risk, 2. Achieve a desired rate of return while minimizing risk, or 3. Maximize returns generated \emph{per} unit risk. Risk is usually measured by the volatility of a portfolio (or asset), which is the variance of its rate of return.
For a given set of assets, this process requires as inputs the rates of returns for each asset, along with their covariances. As the true asset returns are unknown, in practice, these are estimated or forecasted using various techniques that leverage historical data.

This task is then framed as an optimization problem, single or multi-objective, which can be solved in a variety of ways~\cite{cornuejols2006optimization,kalayci2017review,ghahtarani2022robust}.

A typical procedure is to solve it as a convex optimization problem and generate an efficient frontier of portfolios such that no portfolio can be improved without sacrificing some measure of performance (e.g., returns, risk).
Let $w$ be the weight vector for a set of assets, $\mu$ be the expected returns, the portfolio risk can be described as $ w^T \Sigma w $, for covariance matrix $\Sigma$. To achieve a desired rate of return $\mu^*$, we can solve the portfolio optimization problem:
\begin{equation*}
\begin{aligned}
\min_{w} \quad & w^{T}\Sigma w \\
\text{s.t.} \quad & w^{T}\mu \geq \mu^{*} \\
& w_i \geq 0 \\
& \sum_{i} w_i = 1
\end{aligned}
\end{equation*}
Varying $\mu^{*}$ gives us the aforementioned efficient frontier.

Another common objective is the Sharpe Ratio~\cite{sharpe1998sharpe, chen2011all}, which measures the return per unit risk. Formally, for portfolio $p$, the Sharpe Ratio is defined as:

\[ \text{Sharpe Ratio}_{p} = \frac{E[R_{p} - R_{f}]}{\sigma_p} \]

where $R_p$ are the returns of the portfolio, $\sigma_p$ is the standard deviation of these returns, and $R_f$ is a constant risk-free rate (e.g., US Treasuries, approximated by 0.0\% in recent history). Although tricky to solve in its direct form --
\[\max_{w} \frac{\mu^{T}w - R_{f}}{\left(w^{T} \Sigma w\right)^{1/2}}\]
-- it can be framed as a convex optimization problem through the use of a variable substitution~\cite{cornuejols2006optimization}. We choose the Sharpe Ratio as our desired objective function for this study as we can optimize for risk-adjusted returns without having to specify explicit figures for minimum expected returns or maximum risk tolerance.

\subsection{Reinforcement Learning}
Reinforcement Learning (RL) is a sub-field of machine learning that refers to a class of techniques that involve learning by optimizing long-term reward sequences obtained by interactions with an environment~\cite{sutton2018reinforcement}.
An environment is typically formalized by means of a Markov Decision Process (MDP). An MDP consists of a 5-tuple ${ (S,A,P_{a},R_{a},\gamma)}$, where:
\begin{itemize}
  \item $S$ is a set of states
  \item $A$ is a set of actions
  \item $P_{a}(s,s')=\Pr(s_{t+1}=s'\mid s_{t}=s,a_{t}=a)$ is the probability that action $a$ in state $s$ at time $t$ will lead to state $s'$ at time $t+1$
  \item $R_{a}(s,s')$ is the immediate reward received after transitioning from state $s$ to state $s'$, due to action $a$
  \item $\gamma$ is a discount factor between $[0,1]$ that represents the difference in importance between present and future rewards
\end{itemize}

A solution to an MDP is a policy $\pi$ that specifies the action $\pi(s)$ that the decision maker will choose when in state $s$. The objective is to choose a policy $\pi$ that will maximize the expected discounted sum of rewards over a potentially infinite horizon:

\[{E\left[\sum _{t=0}^{\infty }
{\gamma^{t}R_{a_{t}}(s_{t},s_{t+1})}\right]}\]

The field of Deep Reinforcement Learning (DRL) leverages the advancements in Deep Learning by using Neural Networks as function approximators to estimate state-action value functions, or to learn policy mappings $\pi$. These techniques have seen tremendous success in game-playing, robotics, continuous control, and finance~\cite{mnih2013playing,berner2019dota,nguyen2019review,hambly2021recent,charpentier2021reinforcement}.

\subsubsection{RL for Portfolio Allocation}
Given its success in stochastic control problems, RL extends nicely to the problem of portfolio optimization. Therefore, it is not surprising that the use of DRL to perform tasks such as trading and portfolio optimization has received a lot of attention lately.
Recent methods focus on learning deep features and state representations, for example, through the use of embedding features derived from deep neural networks such as autoencoders and LSTM models. These embeddings capture price related features which can range from technical indicators~\cite{wang2019alphastock,soleymani2020financial, Wang_Huang_Tu_Zhang_Xu_2021}, to information extracted from news in order to account for price fluctuations~\cite{ye2020reinforcement}. Other proposed features use attention networks or graph structures~\cite{Wang_Huang_Tu_Zhang_Xu_2021,wang2019alphastock} to perform cross-asset interrelationship feature extraction.

\section{Problem Setup}
We frame the portfolio optimization problem in the RL setting. As described in Section~\ref{sec:background}, RL entails learning in a framework with interactions between an agent and an environment. For the portfolio optimization setting, we create an environment that simulates the US Equities market (using market data replay), and create observation states derived from the observed asset prices. The agent's actions output a set of portfolio weights, which are used to rebalance the portfolio at each timestep.

\subsection{Actions}
For portfolio allocation over $N$ assets, an agent selects portfolio weights $\bm{w} = [{w_1, \dots ,w_N}]$ such that $\sum_{i=1}^{N} w_{i} = 1$, where $0 \le w_i \le 1$.
An asset weight of $0$ indicates zero holdings of a particular asset in a portfolio, whereas a weight of $1$ means the entire portfolio is concentrated in said asset. In extensions of this framework, $w_i < 0$ would allow for shorting an asset, whereas $w_i > 1$ indicates a leveraged position. However, for our case, we restrict actions to non-leveraged long-only positions. These constraints can be enforced by applying the softmax function to an agent's continuous actions.

\subsection{States}\label{sec:States}
An asset's price at time $t$ is denoted by $P_t$. The one-period simple return is defined as $R_t = \frac{P_t - P_{t-1}}{P_{t-1}}$. Consequently, the one-period gross return can be defined as $\frac{P_t}{P_{t-1}} = R_t + 1$. Further, we can define the one-period log return as $r_t = \log\!\left(\frac{P_t}{P_{t-1}}\right) = \log(R_t + 1)$. For our setting, we choose the time period to be daily, and therefore calculate daily log returns using end-of-day close prices.
An asset's log returns over a lookback period $T$ can then be captured as $\bm{r_t} = [{r_{t-1}, r_{t-2},\dots,r_{t-T + 1}}]$. In our case, the lookback period is $T = 60$ days.

For a selection of $n+1$ assets -- $n$ securities and cash (denoted by $c$) -- we form the agent's observation state at time $t$, $\bm{S_t}$ as a $[(n+1) \times T]$ matrix:

\[
{S_t} = \begin{bmatrix}
    w_{1} & r_{1,t-1} &\dots  & r_{1,t-T+1}\\
    w_{2} & r_{2,t-1} & \dots  & r_{2,t-T+1}\\
    \vdots & & \ddots & \vdots\\
    w_{n} &r_{n,t-1} & \dots  & r_{n,t-T+1}\\
    w_{c} & \text{vol}_{20} & \frac{\text{vol}_{20}}{\text{vol}_{60}} & \text{VIX}_t\dots
    \end{bmatrix}
\]

The first column is the agent's portfolio allocation vector $\bm{w}$ as it enters timestep $t$. This might differ slightly from the portfolio weights it chooses at the timestep before, as we convert the continuous weights into an actual allocation (whole shares only), and rebalance the allocation such that it sums to $1$.

For each non-cash asset, we include the log returns over $T$. These are represented by the vector $[{r_{n,t-1}, \dots, r_{n, t-T+1}}]$ for asset $n$ in the state matrix above.

Additionally, in the last row, we include three market volatility indicators at time $t$: $\text{vol}_{20}$, $\frac{\text{vol}_{20}}{\text{vol}_{60}}$, $\text{VIX}$, which we describe in detail in Section~\ref{sec:experiments}.

\subsection{Reward}\label{sec:Reward}
Rather than maximizing returns, most modern portfolio managers attempt to maximize risk-adjusted returns. Since we wish to utilize DRL for portfolio allocation, we want a reward function that helps optimize for risk-adjusted returns. The Sharpe ratio is the most widely-used measure for this, however, it is inappropriate for online learning settings as it is defined over a period of time $T$.
To combat this, we use the Differential Sharpe Ratio $D_t$~\cite{moody1998performance} which represents the risk-adjusted returns at each timestep $t$ and has been found to yield more consistent returns than maximizing profit~\cite{moody2001learning, dempster2006automated}. Therefore, an agent that aims to maximize its future Differential Sharpe rewards learns how to optimize for risk-adjusted returns.

We can define the Sharpe Ratio over a period of $t$ returns $R_t$, in terms of estimates of the first and second moments of the returns' distributions:

\[S_t=\frac{A_t}{K_t(B_t-A_t^2)^{1/2}}\]

with

\[A_t=\frac{1}{t}\sum_{i=1}^t R_i~~\text{and}~B_t=\frac{1}{t}\sum_{i=1}^t R_i^2 ~, ~ K_t=\left(\frac{t}{t-1}\right)^{1/2}\]

where $K_t$ is a normalizing factor.

$A$ and $B$ can be recursively estimated as exponential moving averages of the returns and standard deviation of returns on time scale ${\eta}^{-1}$. We can obtain a \emph{differential} Sharpe ratio $D_t$ by expanding $S_t$ to first order in $\eta$:
\[S_t \approx S_{t-1}+\eta D_t|_{\eta=0} + O(\eta^2)\]
Where Differential Sharpe Ratio $D_t$:

\[D_t\equiv\frac{\partial S_t}{\partial \eta}=\frac{B_{t-1}\Delta A_t-\frac{1}{2}A_{t-1}\Delta B_t}{(B_{t-1}-A_{t-1}^2)^{3/2}}\]

with

\[A_t=A_{t-1}+\eta\Delta A_t\]

\[B_t=B_{t-1}+\eta\Delta B_t\]

\[\Delta A_t= R_t-A_{t-1}\]

\[\Delta B_t=R_t^2-B_{t-1}\]
initialized with $A_0=B_0=0$. We pick $\eta \approx \frac{1}{252}$ (a year has approximately 252 trading days).

\subsection{Learning Algorithm}
RL algorithms can be mainly divided into two categories, model-based and model-free, depending on whether the agent has access to or has to learn a model of the environment.
Model-free algorithms seek to learn the outcomes of their actions through collecting experience via algorithms such as Policy Gradient, Q-Learning, etc. Such an algorithm will try an action multiple times and adjust its policy (its strategy) based on the outcomes of its action in order to optimize rewards.

\subsubsection{Policy Optimization}
Policy optimization methods are centered around the policy $\pi_{\theta}(a|s)$ which is the function that maps the agent's state $s$ to the distribution of its next action $a$. These methods optimize the parameters $\theta$ either by gradient ascent on the performance objective $J(\pi_{\theta})$ or by maximizing local approximations of $J(\pi_{\theta})$. This optimization is almost always performed on-policy since the experiences are collected using the latest learned policy, and then using that experience to improve the policy.
Some examples of popular policy optimization methods are A2C/A3C~\cite{mnih2016asynchronous} and PPO~\cite{schulman2017proximal}. For our experiments we use PPO.

\subsection{RL Environment Specifics} \label{sec:Environment}
The environment serves as a wrapper for the market, sliding over historical data in an approach called market replay. It also serves as a broker and exchange; at every timestep, it processes the agents' actions and rebalances the portfolio using the latest prices and the given allocation. As the day shifts and new prices are received, it communicates these to the agent as observations, along with the Differential Sharpe reward. For the purposes of this study, we assume that there are no transaction costs in the environment, and we allow for immediate rebalancing of the portfolio.

At the beginning of each timestep $t$, the environment calculates the current portfolio value:
\[ \text{port\_val}_t = \sum_{i} P_{i,t} \cdot \text{shares}_{i,t-1} + c_{t-1} \]
In the above expression, $P_i$ is the price of index $i$ at day $t$, $\text{shares}_{i,t-1}$ are the index shares at $t-1$, and $c_{t-1}$ is the amount of cash at $t-1$.

In order to calculate $\text{shares}_{i,t}$ and $c_{t}$, the environment allocates $\text{port\_val}_t$ to the indices and cash according to the new weights $w_i$. Next, it rebalances the portfolio weights $w_i$ to $w_{i\_\text{reb}}$ by multiplying $w_i$ with the current portfolio value, rounding down the number of shares and converting the remaining shares into cash.

After rebalancing, the environment creates the next state $S_{t+1}$ and proceeds to the next timestep $t+1$.
It calculates the new portfolio value based on $P_{t+1}$ and computes the reward $R_t=D_{t}$ which it returns to the agent.

\section{Experiments}\label{sec:experiments}
\subsection{Data \& Features}\label{sec:exp-data}
\begin{figure}[htb]
    \centering
    \includegraphics[width=0.48\textwidth]{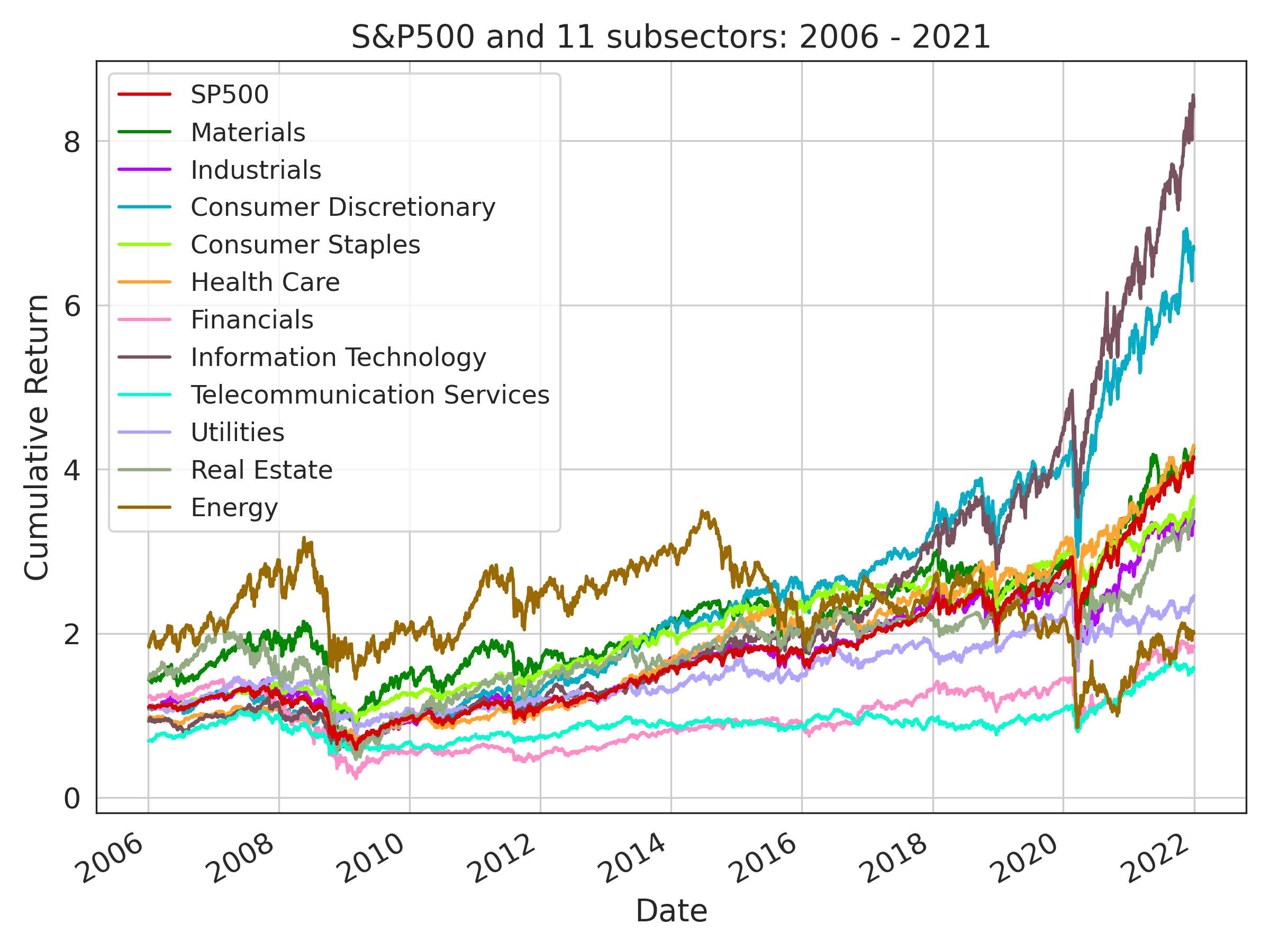}
    \caption{S\&P 500 and its 11 sector indices between 2006 and 2021.}
    \label{fig:port-change}
\end{figure}

For our experiments, we use daily adjusted close price data of the S\&P 500 sector indices as shown in Figure~\ref{fig:port-change}, the VIX index and the S\&P 500 index between $2006$ and $2021$ (inclusive), extracted from Yahoo Finance. The price data is used to compute log returns, as described in Section~\ref{sec:States}.

To capture market regime, we compute three volatility metrics from the S\&P 500 index. The first one, $\text{vol}_{20}$, is the $20$-day rolling window standard deviation of the daily S\&P 500 index returns, the second, $\text{vol}_{60}$, is the $60$-day rolling window standard deviation of the daily S\&P 500 index returns and the third is the ratio of these two $\frac{\text{vol}_{20}}{\text{vol}_{60}}$. This ratio indicates the short-term versus the long-term volatility trend. If $\frac{\text{vol}_{20}}{\text{vol}_{60}} > 1$, that indicates that the past $20$-day daily returns of the S\&P 500 have been more volatile than the past $60$-day daily returns, which might indicate a movement from lower volatility to a higher volatility regime (and vice versa). We use the first and third metrics in the observation matrix, along with the value of the VIX index. These values are standardized by subtracting the mean and dividing by the standard deviation, where the mean and standard deviation are estimated using an expanding lookback window to prevent information leakage.

\subsection{Deep RL Approach}

\subsubsection{Training Process}
Although financial data is notoriously scarce (at least on the daily scale), we want to test the DRL framework across multiple years (backtests). Additionally, financial time series exhibit non-stationarity~\cite{cont2001empirical}; this can be tackled by retraining or fine-tuning models by utilizing the most recently available data. In light of these stylized facts, we devise our experiment framework as follows:

The data is split into $10$ sliding window groups (shifted by $1$-year). Each group contains $7$ years worth of data, the first $5$ years are used for training, the next $1$ year is a burn year used for training validation, and the last year is kept out-of-sample for backtesting.

During the first round of training, we initialize $5$ agents (different seeds) with the hyperparameters described in the following section. All five agents start training on data from $[2006-2011)$ and their performance is periodically evaluated using the validation period $2011$.
At the end of the first round of training, we save the best performing agent (based on highest mean episode validation reward). The final year ($2012$) is kept held-out for backtesting.

This agent is used as a seed policy for the next group of 5 agents in the following training window $[2007-2012)$, validation year $2012$ and testing year $2013$, where this experiment is repeated. This process continues till we reach the final validation period of $2020$, generating a total of $50$ agents ($10$ periods x $5$ agents), and $10$ corresponding backtests (described in a following section).

\subsubsection{PPO Implementation \& Hyperparameters} \label{sec:Hyperparameters}

We use the StableBaselines3~\cite{stable-baselines3} implementation of PPO, and report the hyperparameters used in Table~\ref{table:hyperparameters}. These were picked based on empirical studies~\cite{henderson2018deep, engstrom2019implementation, rao2020make}, as well as a coarse grid search over held-out validation data.

\begin{table}[t]
\centering
\begin{tabular}{l|l}
    training\_timesteps & 7.5M \\
    n\_envs & 10\\
    n\_steps & 756\\
    batch\_size & 1260\\
    n\_epochs & 16\\
    gamma & 0.9\\
    gae\_lambda & 0.9\\
    clip\_range & 0.25\\
    learning\_rate & 3e-4 annealed to 1e-5\\
\end{tabular}
\caption{Hyperparameters used for PPO.}
\label{table:hyperparameters}
\end{table}

Additionally, we make use of the Vectorized SubProcVecEnv environment wrappers provided by StableBaselines3 to collect experience rollouts through multiprocessing across independent instances of our environment. Therefore, instead of training the DRL agent on one environment per step, we train on $n\_envs=10$ environments per step in order to gain more diverse experience and speed up training.

Each round of training lasted a total 7.5M timesteps so as to have approximately 600 episodes per round per environment:
$ (\text{252 trading days per yr} \times \text{5 yrs per round}) \times (\text{10 environments}) \times (\text{600 episodes}) \approx \text{7.5M timesteps} $.
The rollout buffer size was set to $n\_steps = 252 \times 3 \times n\_envs$ so as to collect sufficient experiences across environments. We set up the learning rate as a decaying function of the current progress remaining, starting from $3e-4$, annealed to a final value of $1e-5$. We used a $batch\_size$ of $1260 = (252 \times 5)$, set the number of epochs when optimizing the surrogate loss to $n\_epochs=16$, picked the discount factor $\gamma=0.9$, set the bias-variance trade-off factor for Generalized Advantage Estimator $gae\_lambda=0.9$ and $clip\_range=0.25$.
Additionally, we use a $[64, 64]$ fully-connected architecture with $tanh$ activations, and initialize the policy with a log standard deviation $log\_std\_init=-1$.

\subsection{Mean-Variance Optimization Approach}
As we wish to compare the model-free DRL approach with MVO, we equalize the training and operational conditions. For training, the MVO approach uses a $60$-day lookback period (same as DRL) to estimate the means and covariances of assets. Asset means are simply the sample means over the lookback period. However, we do not directly use the sample covariance, as this has been shown to be subject to estimation error that is incompatible with MVO. To tackle this, we make use of the Ledoit-Wolf Shrinkage operator~\cite{ledoit2004honey}. Additionally, we enforce non-singular and positive-semi-definite conditions on the covariance matrices, setting negative eigenvalues to $0$, and then rebuilding the non-compliant matrices.

Given the estimated means and covariances for a lookback period, we then optimize for the Sharpe Maximization problem and obtain the weights at every timestep. We use the implementation in PyPortfolioOpt~\cite{martin2021pyportfolioopt} to aid us with this process.

\begin{figure*}[htbp]
    \centering
    \includegraphics[width=\textwidth]{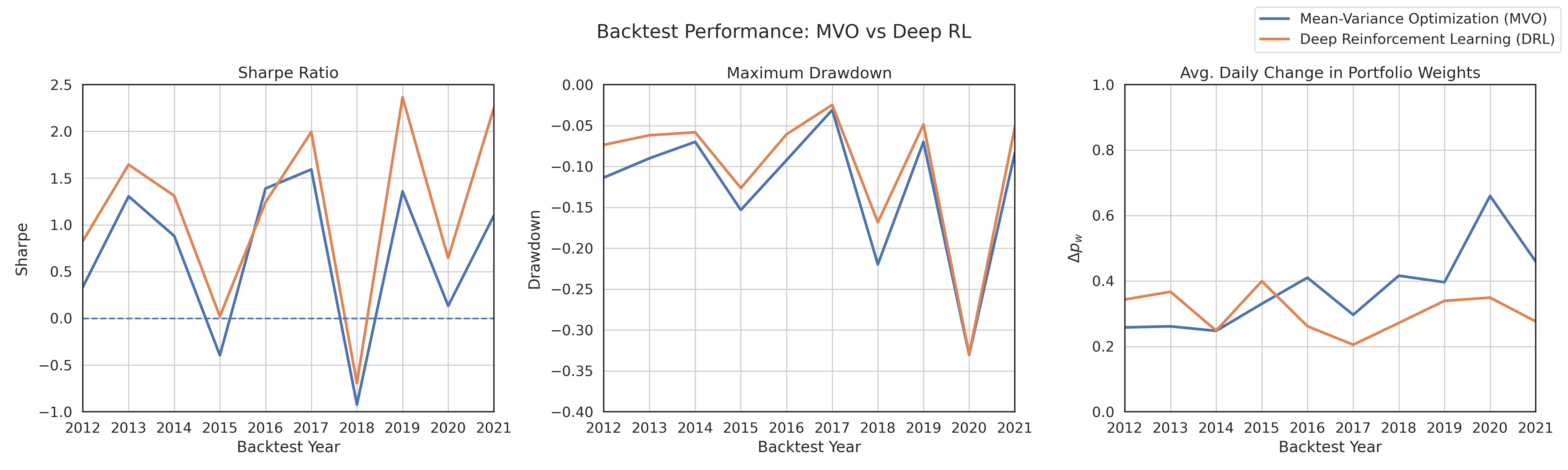}
    \caption{Backtest Results: MVO vs DRL Portfolio Allocation.}
    \label{fig:backtest}
\end{figure*}

\begin{figure*}[htbp]
    \centering
    \includegraphics[width=\textwidth]{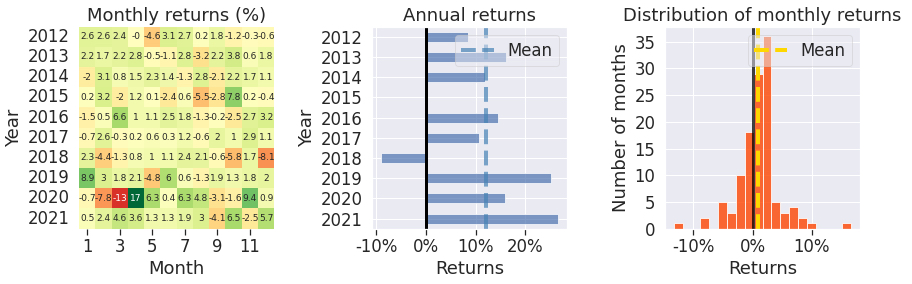}
    \caption{a) DRL Monthly Returns b) DRL Annual Returns c) DRL Monthly Distribution of Returns.}
    \label{fig:DRL_returns}
\end{figure*}

\begin{figure*}[htbp]
    \centering
    \includegraphics[width=\textwidth]{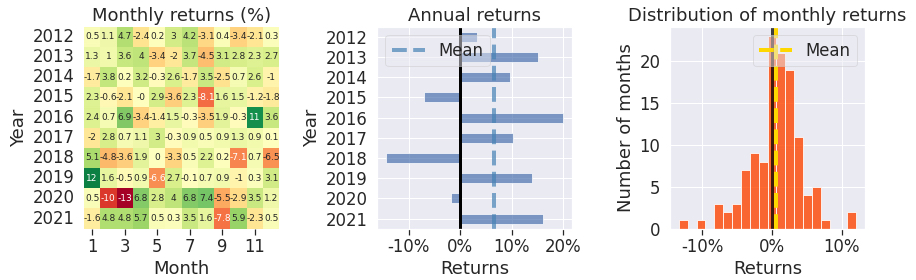}
    \caption{a) MVO Monthly Returns b) MVO Annual Returns c) MVO Monthly Distribution of Returns.}
    \label{fig:MVO_returns}
\end{figure*}

\subsection{Evaluation \& Backtesting}
We evaluate the performance of both techniques through $10$ independent backtests $[2012-2021]$. Both strategies start each backtest period with an all cash portfolio allocation of $\$100,000$. Then, the strategies trade daily using the portfolio weights obtained by each method, enforcing weight constraints $\sum_{i} w_i = 1, 0 \leq w_i \leq 1$, and ensuring only whole number of shares are purchased. By doing so, we can obtain daily portfolio values (and returns), which we subsequently use to compute the statistics we will discuss in the Results section. These are computed with the aid of the Python library Pyfolio.

\textbf{DRL Agent:} We evaluate the trained PPO agents in deterministic mode. For each backtest, the agent used has a gap burn year between the last day seen in training and the backtest period. For example, a DRL backtest carried out in $2012$ would use an agent trained in $[2006-2011)$, with $2011$ being the burn year.

\textbf{MVO:} As the MVO approach does not require any training, it simply uses the past $60$-day lookback period before any given day to calculate portfolio weights. For example, an MVO backtest starting January $2012$ will use data starting October $2011$ (this $60$-day window shifts with each day).

\section{Results}\label{sec:results}

Figure~\ref{fig:backtest} illustrates the performance metrics obtained by applying the aforementioned backtest process on all testing periods [2012-2021]. The DRL agent outperforms the MVO portfolio by exhibiting higher Sharpe and lower yearly maximum drawdowns in virtually every year throughout the backtest period. It also outperforms the MVO portfolio in terms of having marginally lower maximum drawdown.

\begin{table}[t]
    \centering
    \begin{tabular}{|c|c|c|}
    \hline
    \textbf{Metric} & \textbf{DRL} & \textbf{MVO} \\ \hline
    Annual return & 0.1211 & 0.0653 \\ \hline
    Cumulative returns & 0.1195 & 0.0650 \\ \hline
    Annual volatility & 0.1249 & 0.1460 \\ \hline
    Sharpe ratio & 1.1662 & 0.6776 \\ \hline
    Calmar ratio & 2.3133 & 1.1608 \\ \hline
    Stability & 0.6234 & 0.4841 \\ \hline
    Max drawdown & -0.3296 & -0.3303 \\ \hline
    Omega ratio & 1.2360 & 1.1315 \\ \hline
    Sortino ratio & 1.7208 & 1.0060 \\ \hline
    Skew & -0.4063 & -0.3328 \\ \hline
    Kurtosis & 2.7054 & 2.6801 \\ \hline
    Tail ratio & 1.0423 & 0.9448 \\ \hline
    Daily value at risk & -0.0152 & -0.0181 \\ \hline
    \end{tabular}
    \caption{Statistics for the DRL and MVO approaches. All metrics are averaged across $10$ backtests (backtesting period: ${[}2012-2021{]})$, except Max Drawdown which is reported as the maximum seen in any period.}
\label{tab:statistics}
\end{table}

To compare overall performance on the entire backtest period between the two methods, we compute the average performance across all 10 backtest periods. For DRL, we average the performance across the 5 agents (each trained on a different seed) for each year and then average performance across all backtest periods. Similarly, for MVO, we average its performance across all 10 years. By looking at Table~\ref{tab:statistics} we observe that DRL annual returns and Sharpe ratio are ${\approx}1.85\times$ higher than those of the MVO portfolio. The DRL strategy achieves a Sharpe ratio of 1.17 over the full backtest period, compared to 0.68 for MVO.

Figure~\ref{fig:DRL_returns}a) and Figure~\ref{fig:MVO_returns}a) plot the monthly returns over all backtest periods for the two methods. It is evident that DRL is experiencing more steady returns month-to-month than MVO. On the other hand, MVO swings between periods of high returns to periods of low returns a lot more frequently without a steady positive return trajectory. Similarly, in Figure~\ref{fig:DRL_returns}b) and Figure~\ref{fig:MVO_returns}b), we plot the annual returns for the two methods. The vertical dashed line indicates the average annual return across the 10 backtests. For DRL we observe positive returns for almost all backtest years which is a lot more consistent than the behavior of MVO's annual returns. Figure~\ref{fig:DRL_returns}c) and Figure~\ref{fig:MVO_returns}c) plot the distribution of monthly returns averaged across all months. The DRL monthly returns distribution has a lower standard deviation and spread than MVO and a positive mean.

Further, we compute the daily portfolio change for each strategy by measuring the change in its portfolio weights. $\Delta{p_w}$ is the absolute value of the element-wise difference between two allocations (ignoring the cash component). As buying and selling are treated as individual transactions, $\Delta{p_w} \in [0.0, 2.0]$. For example, take a case where the portfolio at time $t-1$ is concentrated in non-cash asset A, and at time $t$ is entirely concentrated in non-cash asset B. This requires selling all holdings of A, and acquiring the equivalent shares in B, leading to $\Delta{p_w} = 2.0$.

Using metric $\Delta{p_w}$, we observe that the Reinforcement Learning strategy has less frequent changes to its portfolio. In practice, this would result in lower average transaction costs. In particular, the average change in portfolio composition is nearly double for Mean-Variance portfolio compared to the DRL strategy during market downturn in March 2020, as shown in Figure~\ref{fig:backtest}, when trading conditions were particularly challenging (i.e. significantly lower market liquidity and elevated bid/ask spreads). Finally, the DRL strategy's performance is derived from the average of five individual agents initialized with different seeds, providing additional regularization which is likely to result in a more stable out-of-sample strategy compared to the MVO strategy.

\section{Conclusion}
We highlight our key contributions as follows:

\begin{itemize}
    \item We have designed a simple environment that serves as a wrapper for the market, sliding over historical data using market replay. The environment can allocate multiple assets and can be easily modified to reflect transaction costs.
    \item We compare our framework's performance during ten backtest experiments over different periods for the US Equities Market using S\&P 500 Sector indices. Our experiments demonstrate the improved performance of the deep reinforcement learning framework over Mean-Variance portfolio optimization.
    \item The profitability of the framework surpasses MVO in terms of Annual Returns, Sharpe ratio and Maximum Drawdown. Additionally, we observe that DRL strategy leads to more consistent returns and more stable portfolios with decreased turnover. This has implications for live-deployment, where transaction costs and slippage affect P\&L.
\end{itemize}

\subsection{Future Work}
In our future work, we would like to model transaction costs and slippage either by explicitly calculating them during asset reallocation or as a penalty term to our reward.
Moreover, we would like to explore adding a drawdown minimization component to our reward that will potentially help the agent learn a more consistent trading strategy.

Another area of exploration is training a regime switching model which will balance its funds amongst two agents depending on market volatility (low vs high). One of them will be a low-volatility trained agent and the other a high volatility trained agent. We would like to compare performance between our current implicit regime parametrization and an explicit one.

\section*{Acknowledgements}
This paper was prepared for information purposes by the Artificial Intelligence Research group of J.~P.~Morgan Chase \& Co.~and its affiliates (``J.~P.~Morgan''), and is not a product of the Research Department of J.~P.~Morgan. J.~P.~Morgan makes no representation and warranty whatsoever and disclaims all liability, for the completeness, accuracy or reliability of the information contained herein.  This document is not intended as investment research or investment advice, or a recommendation, offer or solicitation for the purchase or sale of any security, financial instrument, financial product or service, or to be used in any way for evaluating the merits of participating in any transaction, and shall not constitute a solicitation under any jurisdiction or to any person, if such solicitation under such jurisdiction or to such person would be unlawful.
{\small \bibliography{aaai23}}

\end{document}